\documentclass[prl,aps,twocolumn,reprint,floatfix,superscriptaddress,longbibliography,amssymb]{revtex4-1}
\usepackage[utf8]{inputenc} 
\usepackage{graphicx}
\usepackage{amsmath}
\usepackage{float}

\newcommand{\be}{\begin{equation}}
\newcommand{\ee}{\end{equation}}
\newcommand{\bea}{\begin{eqnarray}}
\newcommand{\eea}{\end{eqnarray}}
\newcommand{\ben}{\begin{equation*}}
\newcommand{\een}{\end{equation*}}
\newcommand{\ba}{\begin{align}}
\newcommand{\ea}{\end{align}}

\newcommand{\mbf}{\mathbf}
\newcommand{\mrm}{\mathrm}
\newcommand{\su}{|\mathord{\uparrow}\rangle}
\newcommand{\sd}{|\mathord{\downarrow}\rangle}
\newcommand{\updownarrows}{\uparrow\mathrel{\mspace{-1mu}}\downarrow}

\begin{document}

\title{Dynamical spin-orbit coupling of a quantum gas}

\author{Ronen M. Kroeze}
\affiliation{Department of Physics, Stanford University, Stanford, CA 94305, USA}
\affiliation{E.~L.~Ginzton Laboratory, Stanford University, Stanford, CA 94305, USA}
\author{Yudan Guo}
\affiliation{Department of Physics, Stanford University, Stanford, CA 94305, USA}
\affiliation{E.~L.~Ginzton Laboratory, Stanford University, Stanford, CA 94305, USA}
\author{Benjamin L. Lev}
\affiliation{Department of Physics, Stanford University, Stanford, CA 94305, USA}
\affiliation{E.~L.~Ginzton Laboratory, Stanford University, Stanford, CA 94305, USA}
\affiliation{Department of Applied Physics, Stanford University, Stanford, CA 94305, USA}

\date{\today}

\begin{abstract}
We realize the \textit{dynamical} 1D spin-orbit-coupling (SOC) of a Bose-Einstein condensate confined within an optical cavity.  The SOC emerges through spin-correlated momentum impulses delivered to the atoms via  Raman transitions. These are effected by classical pump fields acting in concert with the quantum dynamical cavity field. Above a critical pump power, the Raman coupling emerges as the atoms superradiantly populate the cavity mode with photons. Concomitantly, these photons cause a back-action onto the atoms, forcing them to order their spin-spatial state. This SOC-inducing superradiant Dicke phase transition  results in a spinor-helix polariton condensate. We observe emergent SOC through spin-resolved atomic momentum imaging. Dynamical SOC in quantum gas cavity QED, and the extension to dynamical gauge fields, may enable the creation of Meissner-like effects, topological superfluids, and exotic quantum Hall states in coupled light-matter systems.
\end{abstract}

\maketitle

Quantum simulation  in the ultracold atomic physics setting has been enriched by techniques using laser-induced atomic transitions to create synthetic gauge fields~\cite{Spielman2009,Dalibard2011,Goldman2014}, including spin-orbit-coupling (SOC)~\cite{Lin2011}.  Quantum gases in synthetic gauge fields may allow the creation of exotic quantum phases such as  topological superfluids in a pristine environment~\cite{Jiang:2011cw,Goldman2014,Ruhman:2015dk}. At the same time,   strong and tunable atom-atom interactions mediated by  cavity QED light-matter coupling has introduced new capabilities into quantum simulation~\cite{Kimble1998,Ritsch2013,Vaidya:2018fp,Guo:2018wh,Guo:2018tu}.  As such, many-body cavity QED provides unique opportunities for exploring quantum phases and transitions away from equilibrium~\cite{Diehl2010,Sieberer2013,Ritsch2013,Sieberer:2016ej,Kirton:2018vv}. 

Our work combines these two techniques---many-body cavity QED and synthetic gauge fields---for the creation of a novel quantum system exhibiting \textit{dynamical} spin-orbit coupling.  We experimentally demonstrate the emergence of SOC in a Bose-Einstein condensate (BEC) via the use of a cavity field possessing its own quantum dynamics.   Our experiment realizes key aspects of several (previously unrealized) theoretical proposals for creating exotic quantum many-body states via cavity-induced dynamical gauge fields, including SOC~\cite{Mivehvar2014,Deng:2014gqa,Padhi:2014go,Dong:2014cm,Dong2015,Mivehvar2015,Zhu:2016cj,Kollath:2016hs,Sheikhan2016,Zheng:2016fa,Sheikhan2016rc,Halati:2017cu,Halati:2019ij}~\footnote{The ring cavity proposals~\cite{Mivehvar2014,Dong:2014cm,Dong2015,Mivehvar2015,Zhu:2016cj} induce  SOC by directly driving the cavity modes with coherent fields. This fixes the relative phase of the driving fields at any field strength.  By contrast, the Fabry-P\'{e}rot cavity proposals~\cite{Deng:2014gqa,Padhi:2014go,Kollath:2016hs,Sheikhan2016,Zheng:2016fa,Sheikhan2016rc,Halati:2017cu,Halati:2019ij}, including this work, allow the cavity field to be populated from vacuum through a scattering process where phase locking (and hence Raman coupling) emerges dynamically.}. By doing so, this work opens  avenues toward observing exotic phenomena predicted in these works as well as the creation of dynamical gauge fields, complementing recent progress demonstrating density-dependent gauge fields using  optical lattices~\cite{Clark2018,Gorg2018}. Specifically, one might be able to explore unusual nonlinear dynamics~\cite{Dong:2014cm}, novel cooling effects in cavity optomechanics~\cite{Yasir2017}, striped and quantum Hall-like phases~\cite{Mivehvar2014,Deng:2014gqa,Mivehvar2015}, artificial Meissner-like effects~\cite{Ballantine:2017dr,Halati:2019ij}, exotic magnetism~\cite{Padhi:2014go,Mivehvar:2019be}, and topological superradiant states~\cite{Pan:2015ei,Yu2018,Zheng2018}. Adding intracavity optical lattices could create states with directed transport, chiral liquids, and chiral insulators~\cite{Zheng:2016fa,Kollath:2016hs,Sheikhan2016,Sheikhan2016rc,Halati:2019ij}.

\begin{figure}[t!]
\includegraphics[width = 0.49\textwidth]{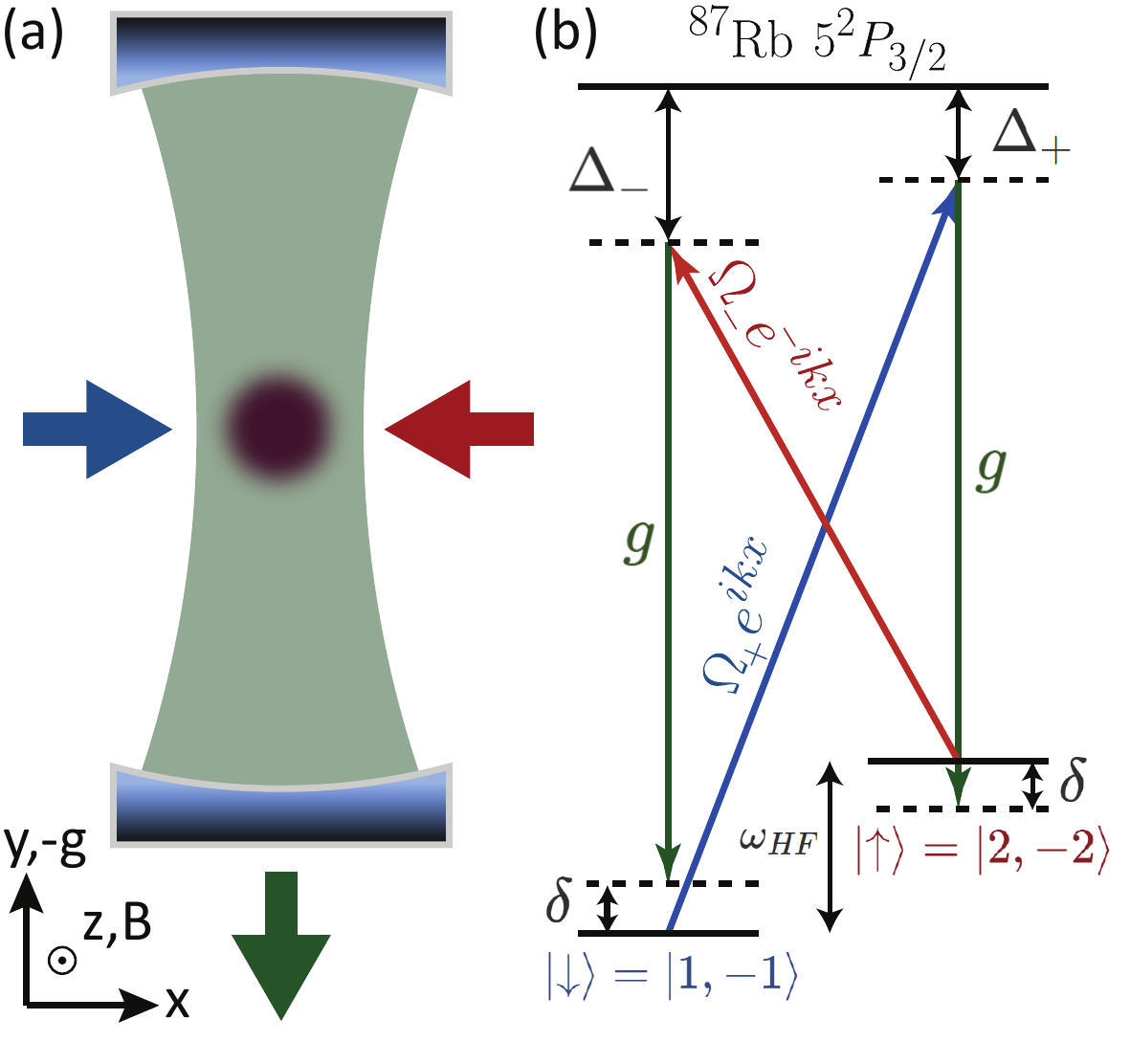}
\caption{(a) Schematic of the experiment. Two Raman pump beams (red and blue arrows), polarized along the cavity axis,  counterpropagate through a BEC of Rb (purple) inside a $\mrm{TEM}_{00}$ cavity. The cavity emission (green arrow) is detected by a single-photon counter, and the atoms are imaged in time-of-flight by a CCD camera (not shown). (b) Level diagram illustrating the cavity-assisted Raman coupling between two hyperfine levels of $^{87}$Rb acting as the spin states. The counterpropagating running-wave nature of the pumps is explicitly notated by $e^{\pm ikx}$. See text for definitions of all quantities.}\vspace{0mm}
\label{experiment}
\end{figure}

Static SOC has been realized in free-space Bose and Fermi quantum gases using two-photon Raman transitions between atomic spin states~\cite{Lin2011,Wu2016,Wang2012,Cheuk2012,Burdick:2016jt}, where the two lasers forming the Raman transition are in classical coherent states with externally fixed intensity. The Raman transition realizes SOC by transferring a recoil momentum to each atom as the spin is flipped, with the recoil direction being correlated with the spin state. The key to our \textit{dynamical} SOC realization is the replacement of one of these classical fields with a cavity mode; see Fig.~\ref{experiment}a and b. Vacuum fluctuations of the cavity mode stimulate  Raman scattering of the pump into this mode. The scattering rate is slow while the atomic spins and positions are disordered.  However, at sufficiently high external pump power, the scattering becomes superradiant due to atomic ordering into a jointly organized spin \textit{and} motional state, reflecting the spin-orbit coupled nature of the system. Because the cavity field feeds back onto the atoms,  the scattering process generating the SOC is dynamical: the SOC depends on the spatial and spin organization of the atoms and vice-versa.  

\begin{figure}[t!]
\includegraphics[width = 0.49\textwidth]{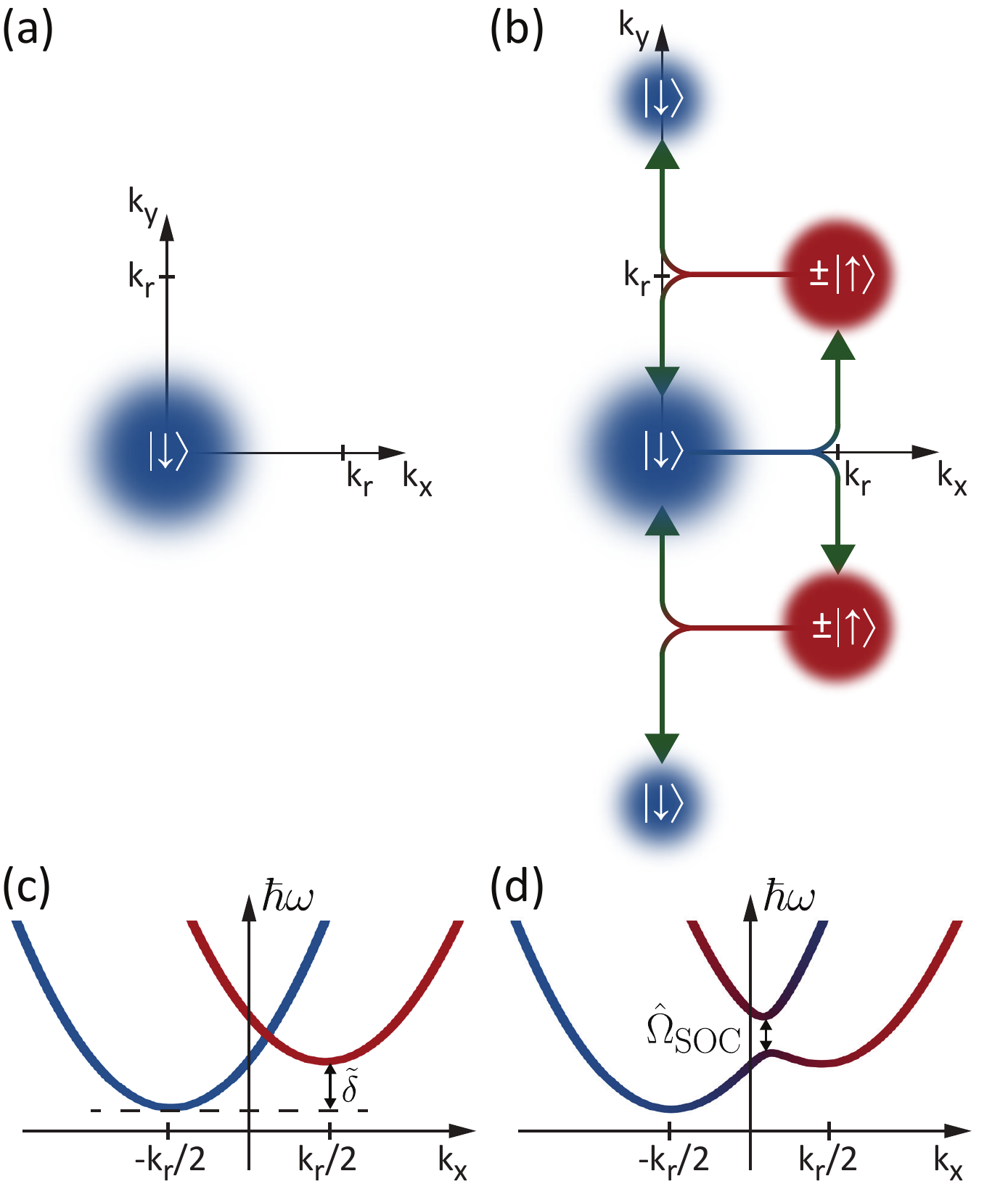}
\caption{(a,b) Momentum-space cartoon of the emergence of SOC. (a) Initially atoms are in a spin-polarized state $\sd$. (b) If the transverse pumping strength is sufficiently strong, SOC emerges and the spin components are in different momentum states. The $\pm$-sign  of the $\su$ spin component indicates the $\mathbb{Z}_2$-symmetry-broken phase freedom. (c,d) Energy-momentum dispersion relation of each spin state, transitioning from free (c) to coupled (d) dispersion bands. The coupling strength $\hat{\Omega}_\mrm{SOC}$ is proportional to $\hat{a}$ and $\hat{a}^\dagger$ and therefore arises dynamically as the atoms scatter pump photons into the cavity. The zero of the momentum has been shifted with respect to the lab frame by $-k_r/2$ in the plot; cf.~the unitary transformations in Eqs.~\ref{Utransform}. }\vspace{0mm} \label{kspace}
\end{figure}

In contrast to systems with standing-wave pump fields in which no SOC arises~\footnote{For standing-wave pumps, the resulting state of spinful atoms in a BEC has been experimentally shown to be a `spinor polariton condensate,' i.e., a jointly organized spin \textit{and} spatial light-matter state of macroscopic population~\cite{Kroeze:2018wd}.  While the $k=0$ ground state of the BEC is necessary for macroscopic population of the ground and first excited momentum states, the broken $\mathbb{U}(1)$ gauge symmetry of the BEC is not relevant.}, SOC emerges at the transition threshold when running-wave fields are used as pumps.  This is because the running-wave pumps,  in conjunction with the cavity mode, impart momentum kicks to the atoms as they flip the atomic spins~\footnote{Checkerboard density wave organization also arises in running-wave-pumped cavities~\cite{Arnold2012}.}.  Momentum is transferred only along the pump axis because the the cavity field is a standing wave. Figure~\ref{kspace} depicts the emergence of SOC, both in terms of occupation of momentum states and in the coupling between the bands. The phase transition results in a spinor-helix-like state where the spin state rotates along $\hat{x}$ with a period commensurate with the pump wavelength.  While the total density remains translationally invariant along $\hat{x}$, both  spin and density are modulated along the cavity axis, as described below.

This dynamical SOC may also be understood from the perspective of cavity-field phase fluctuations. Below threshold, the scattering into the cavity is due to the pump light coupling to incoherent atomic spin and density  wave fluctuations~\cite{Kroeze:2018wd}.  These spinor density-wave fluctuations cause the resonating light  to possess a phase that is both uncorrelated and time-varying with respect to that of the pump field. Therefore, coherent Raman transitions---and thus, SOC---are suppressed due to the random diffusion of the relative phase between the pump and cavity fields.  

Stable SOC emerges only once the phase of the cavity field  locks with respect to the pump fields.  This locking occurs when the pump power reaches a threshold for triggering a nonequilibrium (Hepp-Lieb) Dicke superradiant phase transition~\cite{Ritsch2013,Kirton:2018vv}.  At threshold, the atomic spinor state condenses into helical patterns oriented along the pump axis $\hat{x}$.  There is a helix at each antinode of the cavity field along $\hat{y}$ and the phase of neighboring helices differ by $\pi$; the resultant state is $\left|\psi_\text{helix}\right\rangle=\sd\pm e^{ikx}\cos{ky}\,\su$~\footnote{Reference~\cite{Mivehvar:2019be} also proposed the existence of such a state.}. The broken $\mathbb{Z}_2$ symmetry of the phase transition is reflected in the spontaneous choice of the $\pm$ sign, which determines the helix phase (0 or $\pi$) with respect to the phase of the pump fields.  The helix pattern serves as a grating for the  Bragg-diffraction (i.e., superradiant scattering) of pump photons into the cavity mode.  Superradiance increases the coherent field of the cavity by a factor proportional to the number of atoms.  Moreover, it locks the cavity phase to either 0 or $\pi$ with respect to the phase of the Raman lasers.  This phase choice is correlated with the $\pm$-sign choice in $\left|\psi_\text{helix}\right\rangle$.  

The experiment employs two counterpropagating pump beams with amplitudes $\Omega_+$ and $\Omega_-$ to couple two internal states $|F,m_F\rangle=|1,-1\rangle\equiv\sd$ and $|F,m_F\rangle=|2,-2\rangle\equiv\su$ of a $^{87}$Rb BEC.  This is illustrated in Figs.~\ref{experiment}a and b.  The fields induce two cavity-assisted Raman processes that together generate the Hamiltonian $\hat{H}=-\Delta_c\hat{a}^\dagger\hat{a}+\int\hat{\psi}(\mbf{r})^\dagger\hat{H}_\text{SOC}\hat{\psi}(\mbf{r})d^3\mbf{r}$.  Here, $\Delta_c$ is the cavity detuning, $\hat{a}$ ($\hat{a}^\dagger$) is the annihilation (creation) operator for the intracavity field, and $\hat{\psi}(\mbf{r})=\left[\hat{\psi}_\uparrow(\mbf{r}),\hat{\psi}_\downarrow(\mbf{r})\right]^\top$ is a spinor containing the atomic annihilation operators $\hat{\psi}_{\updownarrows}$.  The SOC Hamiltonian is 
\be
H_{\mrm{SOC}}=\begin{bmatrix}\frac{(\hat{\mbf{p}}+k_r/2\mbf{e}_x)^2}{2m}+D_+-\tilde{\delta} & \hat\Omega_\mrm{SOC}\cos{k_ry} \\
H.c. & \frac{(\hat{\mbf{p}}-k_r/2\mbf{e}_x)^2}{2m}+D_-\end{bmatrix}
\ee
where $k_r$ is the recoil momentum of the transverse pumps, $\mbf{e}_x$ is the unit vector in  $\hat{x}$, $\tilde{\delta}$ is the effective two-level spin splitting set by the Raman detuning $\delta$  minus the (small) AC light shift, and $D_\pm=\frac{g^2(x,z)}{\Delta_\pm}\cos^2(k_ry)\hat{a}^\dagger\hat{a}$ is the dispersive shift~\footnote{Since the differential dispersive shift is small compared to the other terms, we will neglect it from here forward.}. The dynamical Raman coupling strength is
\be
\hat\Omega_\mrm{SOC}=\frac{g(x,z)\Omega_+}{\Delta_+}\hat{a}^\dagger+\frac{g(x,z)\Omega_-}{\Delta_-}\hat{a},
\ee
where $\Delta_\pm=-150$~GHz~$\pm\,\,\omega_\text{HF}$ is the  atomic detuning for each pump, $\omega_\text{HF}=6.8$~GHz the total hyperfine and Zeeman splitting between the two spin states, and $g(x,z)$ the spatially dependent single-atom atom-cavity coupling strength. See supplemental material for a derivation of this SOC Hamiltonian model and its mapping to the Dicke model.  This model is similar to that considered in the recent proposal paper~\cite{Halati:2019ij}, where exotic Meissner-like effects were predicted to exist, as also discussed in Ref.~\cite{Ballantine:2017dr}.  Another recent proposal paper considered a similar Raman coupling scheme in the context of generating exotic spin Hamiltonians~\cite{Mivehvar:2019be}.

\begin{figure}[t!]
\includegraphics[width = 0.49\textwidth]{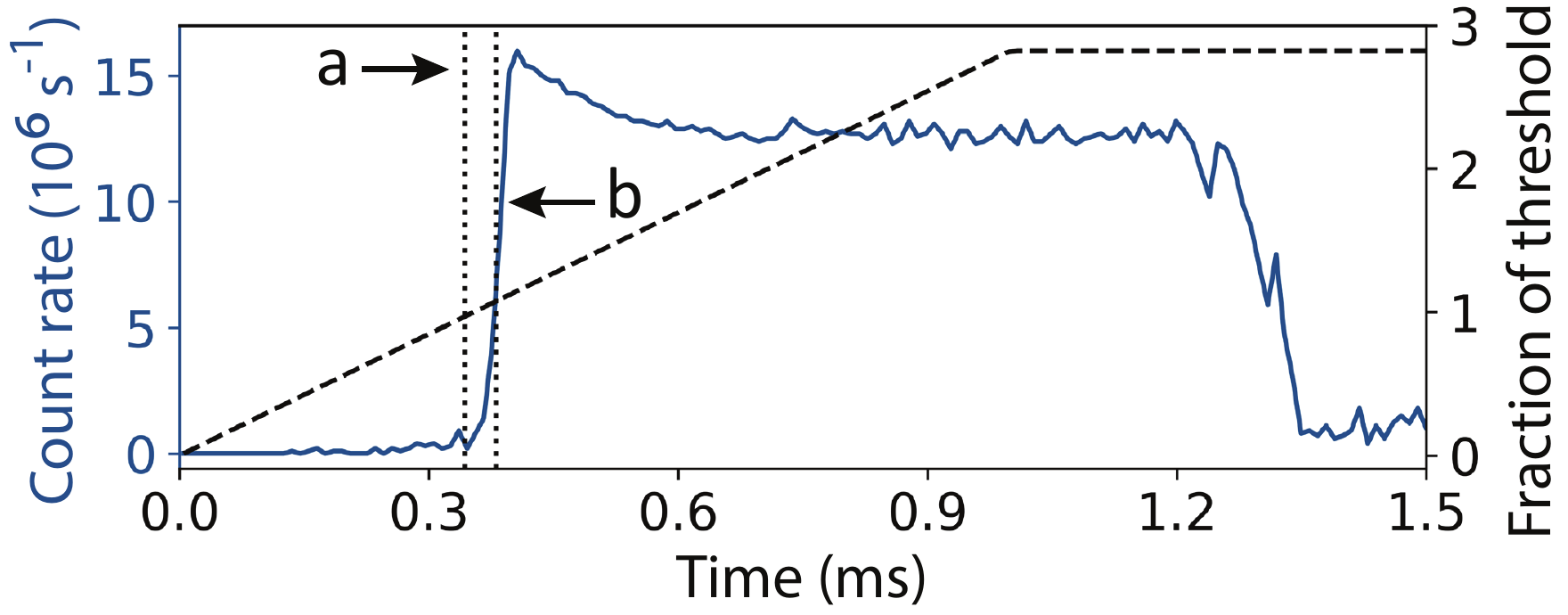}
\caption{Cavity emission detected by single-photon counters (solid blue line), and optical power in the Raman beams (dashed black line), both as function of time. The observed emission is indicative of steady-state SOC up to the $\sim$ms timescale. Labels ``a" and ``b" indicate just before and after the superradiant transition threshold, respectively.} 
\label{00superradiance}\vspace{4mm}
\end{figure}

The SOC arises in this model because each spin state is  addressed by only one of the two Raman processes. For instance, an atom in $\sd$ can only scatter photons into the cavity from $\Omega_+$, since $\Omega_-$ is off-resonance by $\approx2\omega_\text{HF}$. Due to the running-wave nature of the transverse pumps, each scattering event imparts a net momentum along $+\mbf{e}_x$ onto the atom, because the accompanying momentum change $\pm k_r$ along the cavity direction averages to zero since either direction is equally probable. Likewise, an atom originating in $\su$ will receive a net momentum kick along $-\mbf{e}_x$, where the direction is opposite due to the counterpropagating orientation of the running-wave pump beams.  The result---opposite spin states moving in opposite directions---thus realizes SOC. Note however, that the Raman coupling term $\hat\Omega_\mrm{SOC}$ contains the cavity field operators $\hat{a}$ and $\hat{a}^\dagger$. Since the cavity field is determined self-consistently by the dynamics of the atom-spin-cavity system, and is initially in a vacuum state, the SOC term  emerges \textit{dynamically} as the atoms organize to scatter superradiantly.

We now present data demonstrating emergent SOC. A BEC of $4.1(3)\times10^5$ $^{87}$Rb atoms, all prepared in $\sd$, is placed at the center of a $\mrm{TEM}_{00}$ cavity by an optical dipole trap; see Refs.~\cite{Kollar2015,Kroeze:2018wd,Guo:2018wh} for details.  The cavity and pump fields are tuned such that $\Delta_c=-6$~MHz and $\tilde\delta=-10$~kHz. We record the light emitted from the cavity on a single-photon counter. The power of the transverse pumps is gradually increased and then held constant, shown by the black dashed line in Fig.~\ref{00superradiance}. The recorded cavity emission is shown in blue, and rapidly increases when the optical power reaches threshold, indicating the emergence of superradiant scattering and, consequently, the nonzero Raman coupling needed for SOC~\footnote{We note that the duration of emission is longer than that from a single spin-flip process~\cite{Zhiqiang2018}, implying steady-state SOC instead of transient effects.}. The superradiance lifetime is presumably limited by the dephasing of the two pumping beams, which we independently verified is also on the millisecond timescale.

\begin{figure}[t!]
\includegraphics[width = 0.49\textwidth]{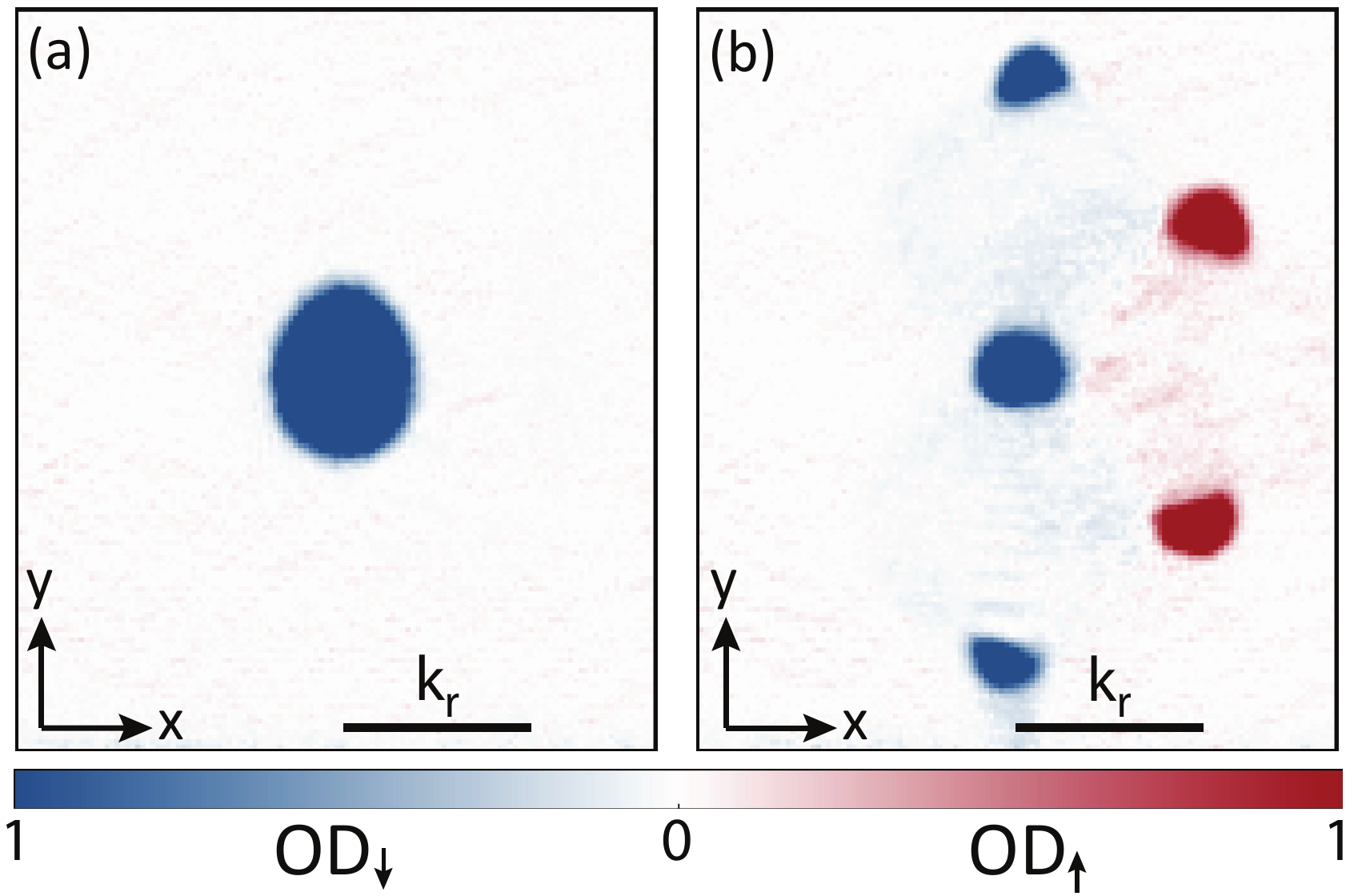}
\caption{Spin-resolved momentum distribution in time-of-flight.  (a) Just below threshold (indicated by the label ``a" in Fig.~\ref{00superradiance}) all atoms are still in $\sd$. (c) Above threshold (indicated by ``b" in Fig.~\ref{00superradiance}), spin-up atoms have acquired a net momentum in the $x$ direction, as shown by the spin-colored Bragg peaks at nonzero momentum. Also shown are second-order diffraction peaks along the cavity direction due to the reverse Raman process.} 
\label{momentum}\vspace{4mm}
\end{figure}

We have observed that the SOC-induced Bragg peaks emerge at the same pump power as the threshold for superradiant cavity emission, as expected; see Fig.~\ref{kspace}. This is determined  by correlating the cavity emission signal in Fig.~\ref{00superradiance} with the spin-resolved, time-of-flight imaging of the atomic gas in Figs.~\ref{momentum}a and b~\footnote{See Ref.~\cite{Kroeze:2018wd} for how spin-resolved imaging is performed in our apparatus}.  These images provide full information about each spin-species' momentum distribution.  Below threshold, the cavity emission is weak and  all atoms are in the initial state, i.e., a zero-momentum spin-polarized state. This is shown in Fig.~\ref{momentum}a. At a time shortly after reaching threshold, a fraction of the atoms have undergone a spin-flip and have scattered  into the two Bragg peaks at $(-k_r,\pm k_r)$, as shown in Fig.~\ref{momentum}b. Additionally, the reverse process occurs, mediated by $\Omega_-$, and repopulates the zero-momentum component as well as scattering  some atoms into $(0,\pm2k_r)$~\footnote{These secondary diffraction peaks do not occur in a single spin-flip process.}. Crucially, the spin-species have now separated in momentum space, with a net difference in momentum component along $\mbf{e}_x$. This is evidence for the SOC state, and the observed momentum distribution indicates that the spin distribution (up to global phase factors) corresponds to the aforementioned spinor-helix state $\left|\psi_\text{helix}\right\rangle$. This state possesses similarities to the persistent spin-helix state observed in semiconductors~\cite{Bernevig:2006gg,Koralek2009} and could be extended to Abelian or non-Abelian `Majorana' spinor-helix states through the use of high-spin lanthanide atoms such as dysprosium (on timescales less than that set by dipolar relaxation)~\cite{Lian:2012eo,Cui:2013ki,Burdick:2015bx,Burdick:2016jt}.  The limited superradiance lifetime hampered our ability to measure both the excitation spectrum of the spinor-helix mode and the position of the SOC band minima in Fig.~\ref{kspace}(d) versus the emergent Raman coupling strength.  Future improvements to the Raman laser lock should improve this lifetime and enable these measurements.

In conclusion, we have observed spin-orbit coupling that emerges through a process of spin-spatial (spinor) self-organization.  This organization arises due to the scattering of running-wave pump fields into the cavity field. Quantum fluctuations of the cavity field stimulate this scattering process, generating a cavity field incoherent with the pump field.  At higher pump power, a runaway self-organization transition induces the superradiant scattering of a field whose phase is locked with the pumps.  The resulting coherent Raman coupling---arising from the mutually coherent pump and cavity fields---induces \textit{dynamical} SOC.   Moreover, the BEC-cavity QED system is strongly coupled and therefore quantum fluctuations can play a role in the SOC dynamics. This is because the spin-spatial self-organization  takes place at an SOC threshold corresponding to only a few  cavity photons wherein quantum fluctuations are non-negligible.  Consequences of this will be explored in future work.

The addition of dynamical SOC to the toolbox of quantum simulation in the nonequilibrium context opens new avenues for the exploration of a wide range of phenomena in quantum gases, e.g., topological superradiant superfluids. Moreover, dynamical artificial gauge fields can be created by a simple modification of the present experiment.  Specifically, by using a multimode cavity (possible with our present apparatus~\cite{Kollar2015}) and by choosing the pump laser frequencies to enhance the effects of their differential dispersive light shift on the spin states, Meissner-like effects can be observed~\cite{Ballantine:2017dr}.  We speculate that with dynamical gauge fields, combined with the strong, sign-changing, and tunable-range photon-mediated interactions provided by multimode cavities~\cite{Vaidya:2018fp,Guo:2018wh,Guo:2018tu}, quantum simulators will be able to create a wide variety of exotic, nonequilibrium quantum matter.

We thank Jonathan Keeling and Sarang Gopalakrishnan for helpful discussions.  We are grateful for funding support from the Army Research Office. 

\section{Supplemental Material}

\subsection{Cavity-mediated spin-orbit coupling}\label{cavitySOC}
For a pair of counterpropagating running-wave Raman lasers with amplitudes  $\Omega_\pm$, phases $\phi_\pm$, and wavenumber $k_r=\frac{2\pi}{\lambda}$, the terms of the total Hamiltonian for the system, including the cavity-assisted Raman coupling between the two spin states $\hat{\psi}_{\uparrow}$ and $\hat{\psi}_{\downarrow}$, are given by
\begin{widetext}
\begin{align}
H_{\uparrow} &= \int d^3 \mbf{r}\, \hat{\psi}^{\dagger}_{\uparrow}(\mbf{r}) \Big[ \frac{\hat{\mbf{p}}^2}{2 m}  + \omega_S + \frac{g^2 (x,z)}{\Delta_+} \cos^2(k_r y) \hat{a}^{\dagger} \hat{a} - \delta \Big] \hat{\psi}_{\uparrow} (\mbf{r}) \nonumber \\
H_{\downarrow} &= \int d^3 \mbf{r}\, \hat{\psi}^{\dagger}_{\downarrow}(\mbf{r}) \Big[ \frac{\hat{\mbf{p}}^2}{2 m}  + \frac{g^2 (x,z)}{\Delta_-} \cos^2(k_r y) \hat{a}^{\dagger} \hat{a} \Big] \hat{\psi}_{\downarrow} (\mbf{r}) \nonumber \\
H_{\mathrm{cavity}} &= -\Delta_c \hat{a}^{\dagger} \hat{a} \nonumber \\
H_{\mrm{Raman}} &= \int d^3 \mbf{r} \left[\frac{g(x,z) \Omega_+ e^{i (k_rx + \phi_+)}}{\Delta_+} \hat{\psi}^{\dagger}_{\uparrow} (\mbf{r}) \hat{\psi}_{\downarrow} (\mbf{r}) \hat{a}^{\dagger} + \frac{g(x,z) \Omega_- e^{i (-k_rx + \phi_-)}}{\Delta_-} \hat{\psi}_\downarrow^{\dagger}(\mbf{r}) \hat{\psi}_\uparrow(\mbf{r}) \hat{a}^{\dagger} + h.c.\right]\cos(k y).
\end{align}
\end{widetext}
Here, $g=g(x,z)\cos(k_r y)$ is an explicitly notated version of the $y$-dependence of the atom-cavity coupling strength, and 
\bea
\omega_S = &&\left[ \frac{\Omega^2_{+}}{6(\Delta_+ + \omega_\text{HF})} +  \frac{\Omega^2_{-}}{6\Delta_-} \right] \nonumber\\
&&- \left[ \frac{\Omega^2_{+}}{6\Delta_+} +  \frac{\Omega^2_{-}}{6(\Delta_- - \omega_\text{HF})} \right]
\eea
is the differential Stark shift due to the two pump beams, with $\omega_\text{HF}$ the energy splitting between the two spin states. Note that we have defined the cavity axis along $\hat{y}$ such that $\hat{z}$ is the quantization axis defined by the magnetic field that we apply and $\hat{x}$ is the direction of the Raman beams. To write the Hamiltonian in the form of a familiar spin-orbit coupling Hamiltonian, we apply the unitary transformations
\begin{align}
\psi_\uparrow &\rightarrow \psi_\uparrow e^{i (k_r x + \phi)/2} \nonumber\\
\psi_\downarrow &\rightarrow \psi_\downarrow e^{-i (k_r x + \phi)/2} \nonumber\\
a &\rightarrow e^{i\Phi} a,\label{Utransform}
\end{align}
where $\phi = (\phi_+-\phi_-)/2$ and $\Phi = (\phi_++\phi_-)/2$. After this transformation, the different parts of the Hamiltonian become
\begin{widetext}
\begin{align}
H_{\uparrow} &= \int d^3 \mbf{r}\, \hat{\psi}^{\dagger}_{\uparrow}(\mbf{r}) \Big[ \frac{(\hat{\mbf{p}}+k_r\mbf{e}_x/2)^2}{2 m}  + \omega_S + \frac{g^2 (x,z)}{\Delta_+} \cos^2(k_r y) \hat{a}^{\dagger} \hat{a} - \delta \Big] \hat{\psi}_{\uparrow} (\mbf{r}) \nonumber \\
H_{\downarrow} &= \int d^3 \mbf{r}\, \hat{\psi}^{\dagger}_{\downarrow}(\mbf{r}) \Big[ \frac{(\hat{\mbf{p}}-k_r\mbf{e}_x/2)^2}{2 m}  + \frac{g^2 (x,z)}{\Delta_-} \cos^2(k_r y) \hat{a}^{\dagger} \hat{a} \Big] \hat{\psi}_{\downarrow} (\mbf{r}) \nonumber \\
H_{\mathrm{cavity}} &= -\Delta_c \hat{a}^{\dagger} \hat{a} \nonumber \\
H_{\mrm{Raman}} &= \int d^3 \mbf{r} \left[\hat{\psi}^{\dagger}_{\uparrow} (\mbf{r}) \hat{\psi}_{\downarrow} (\mbf{r})\left(\frac{g(x,z) \Omega_+}{\Delta_+}  \hat{a}^{\dagger} + \frac{g(x,z) \Omega_- }{\Delta_-} \hat{a}\right) + h.c.\right]\cos(k_r y).
\end{align}
\end{widetext}
This Hamiltonian exhibits a typical form of spin-orbit coupling~\cite{Spielman2009,Dalibard2011,Goldman2014}, since the kinetic energy term is modified differently for each spin species.

\subsection{Mapping to the Dicke model}\label{dicke}
To make a connection with existing literature regarding transversely pumped ultracold gases in a cavity, the above Hamiltonian can also be mapped onto the superradiant (Hepp-Lieb) Dicke model~\cite{Ritsch2013,Kirton:2018vv} using the single-recoil approximation. In the lab frame, this corresponds to $\psi_\downarrow = c_\downarrow\psi_0$ and $\psi_\uparrow = c_\uparrow\psi_1$, with
\begin{align}
\psi_0 &=1\nonumber\\
\psi_1 &= \sqrt{2}e^{ik_rx}\cos{k_ry}.
\end{align}
After the unitary transformation in Eq.~\ref{Utransform}, these become
$\psi_0 =e^{ik_rx/2}$ and $\psi_1 = \sqrt{2}e^{ik_rx/2}\cos(k_ry)$. Inserting these into the above equations, and evaluating the integrals, simplifies the Hamiltonian to
\begin{align}
H = &(2\omega_r+\omega_S-\delta)c_\uparrow^\dagger c_\uparrow+\frac{3g_0^2}{4\Delta_+}c_\uparrow^\dagger c_\uparrow a^\dagger a + \frac{g_0^2}{2\Delta_-}c_\downarrow^\dagger c_\downarrow a^\dagger a\nonumber\\&-\Delta_c a^\dagger a +\left(\frac{g_0\Omega_+}{\sqrt{2}\Delta_+}c_\uparrow^\dagger c_\downarrow+\frac{g_0\Omega_-}{\sqrt{2}\Delta_-}c_\downarrow^\dagger c_\uparrow\right)(a+a^\dagger),
\end{align}
where $\omega_r={k_r^2}/{2m}$ is the recoil energy. Taking the Raman couplings to be equal,
\be
\eta_D\equiv\frac{g_0 \Omega_+}{\Delta_+}\sqrt{\frac{N}{2}} = \frac{g_0 \Omega_-}{\Delta_-}\sqrt{\frac{N}{2}},
\ee
and defining the spin-operators as
\begin{align}
\hat{J}_{z} &= \frac{1}{2} (\hat{c}^{\dagger}_{\uparrow} \hat{c}_{\uparrow} - \hat{c}^{\dagger}_{\downarrow} \hat{c}_{\downarrow}) \nonumber \\
\hat{J}_+ &= \hat{c}^{\dagger}_{\uparrow} \hat{c}_{\downarrow} \nonumber \\
\hat{J}_- &= \hat{c}^{\dagger}_{\downarrow} \hat{c}_{\uparrow},
\end{align}
 this Hamiltonian  becomes 
\begin{align}
H =&\left(-\Delta_c + \frac{N g_0^2}{2 \Delta_-} \right) \hat{a}^{\dagger} \hat{a} + (2 \omega_r + \omega_S - \delta) \hat{J}_{z} \nonumber \\
&+ \frac{\eta_D}{\sqrt{N}}(\hat{J}_+ + \hat{J}_-)(\hat{a}^{\dagger} + \hat{a}).
\end{align}
We have assumed here that the normalization condition $\hat{c}^{\dagger}_{\uparrow} \hat{c}_{\uparrow} + \hat{c}^{\dagger}_{\downarrow} \hat{c}_{\downarrow}=N$, where $N$ denotes the total number of atoms, and  we have discarded a constant energy offset and a term $\propto\hat{c}^{\dagger}_{\uparrow}\hat{c}_{\uparrow} $, which is small in the single-recoil limit. The Hamiltonian therefore realizes the  Dicke model exhibiting a superradiant phase transition~\cite{Ritsch2013,Kirton:2018vv}.


%

\end{document}